\documentclass[aps,amsmath,amssymb,amsfonts,
floatfix,showpacs,prb,reprint,twoside,superscriptaddress,linenum]{revtex4-1}

\usepackage{bm}
\usepackage{graphicx}
\usepackage[varg]{txfonts}
\usepackage{lineno}

\renewcommand{\vec}[1]{\bm{#1}}%

\newcommand{\therm}[3]{\left( 
                    \frac{\partial {#1}}{\partial {#2}} 
                        \right)_{#3}}

\begin{document}

\title{Coherent heat-wave generation in SrTiO$_3$ using 
stimulated light-scattering} 
\author{Akitoshi Koreeda, Hideaki Oe, Takahiro Okada, and
Yasuhiro Fujii}

%
\affiliation{Department of Physical Sciences, 
Ritsumeikan University, Kusatsu, 525-8577, Shiga, Japan}

\date{\today}

\begin{abstract}

 Wave propagation of heat in solids, which is often called
  ``second sound'', has attracted increasing attention 
 to its various potential applications such as in thermal
 diodes and  waveguides.
%
%
 It has been suggested that ferroelectrics
 are strong candidates for thermal-wave
 media\cite{Gurevich-Tagantsev}, and several reports on an
 incipient ferroelectric, strontium titanate (SrTiO$_3$), have shown
 a novel wave-like excitation in light-scattering
 spectra\cite{Hehlen, Koreeda-PRL-2007}. Second sound has
 been proposed as the most plausible origin for this excitation
 based on rather indirect demonstrations\cite{Koreeda-PRL-2007,
 Martelli2018}. However, the lack of a direct demonstration still
 leaves controversy\cite{Huberman2019} on the existence of second
 sound in SrTiO$_3$.
 Here, we present a direct demonstration of an
 optically generated coherent heat wave, i.e, coherent second
 sound, in bulk single crystals of SrTiO$_3$.
 The generated wave-excitation has a linear dispersion relation
 with the predicted velocity of second sound in SrTiO$_3$, and
 is overdamped at high temperatures or low frequencies,
 exhibiting apparently diffusive behaviour with the decay rates
 defined by the known values of the thermal diffusivity of
 SrTiO$_3$.
 This work offers not only convincing evidence for the
 existence of second sound in SrTiO$_3$, but also a first
 demonstration of coherent generation and wave-control of
 gigahertz temperature fields in a 3D bulk material.
 As high-quality SrTiO$_3$ is widely available, and it is also known
 as a mother material of superconductors, the coherent excitation
 of temperature wave in this material offers a pathway for
 potential novel wave-based applications in heat management such
 as construction of thermal waveguides and directional devices 
 or heat-control of superconductivity.

\end{abstract} 

\maketitle

Thermal transport in insulators is realised by lattice
vibrations, i.e., phonons, that are quantised sound
waves. Recently, phonon-related physics, such as the phonon Hall
effect\cite{Li2020}, phonon spin/orbital angular
momenta\cite{Ishito2021, Oishi2022}, and phonon
hydrodynamics\cite{Ghosh2022a} has attracted much attention.
Hydrodynamic phonon excitation allows for collective wave
propagation of constituent phonons, i.e., wave propagation of the
temperature field in solids. Such a wave excitation of phonons is
referred to as `second sound', owing to its slower propagation
speed than ordinary (first) sound, although second sound is not
technically a `sound'.  Second sound is a very rare phenomenon,
and only a limited number of materials have been reported as
candidates for its medium\cite{Ackerman-PRL-1969, SS-in-Bi,
Jackson1970, Koreeda-PRL-2007, Huberman2019, Ding2022a}, so
far. Among these materials, strontium titanate (SrTiO$_3$) has
long been a subject of controversy with respect to the existence
of second-sound excitation within it\cite{Hehlen, Scott-TPDS,
Courtens-to-Scott, Yamaguchi-prb-2002, Scott-myth,
Koreeda-PRL-2007, Koreeda-PRB-2009, Scott2013, Nakamura2015,
Huberman2019}.

\begin{figure*}[t]
 \centering \includegraphics[width=0.9\textwidth]{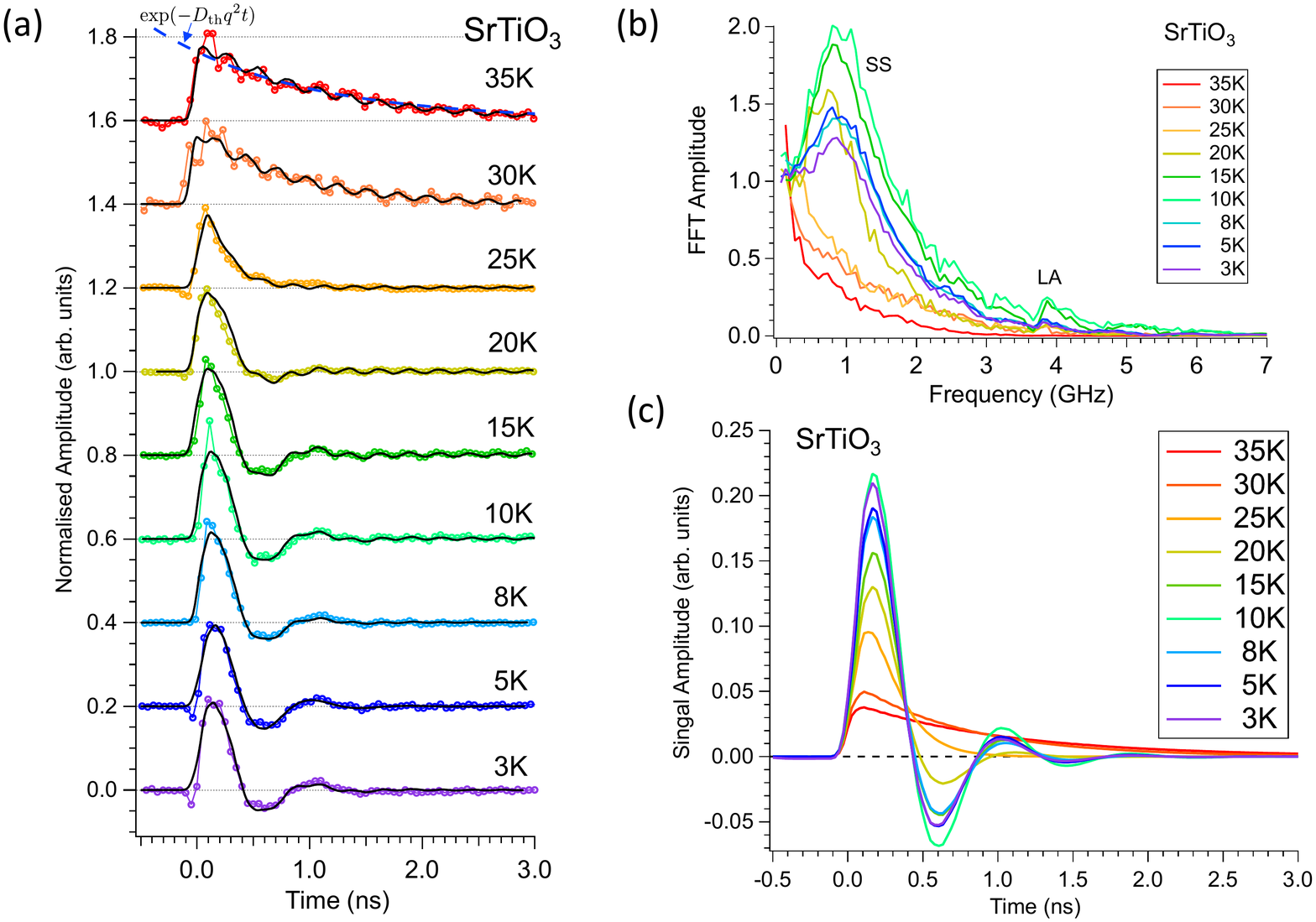}
 \caption{(a) Temperature variations of the temporal profiles
 observed in SrTiO$_3$. All traces were obtained after smoothing
 and subtraction of the residual constant offset due to the
 electronic excited state (see Extended Data Figure 3). (b)
 Fourier-transformed spectra of the observed temporal signals. SS
 and LA denote second sound and longitudinal acoustic phonon,
 respectively.  (c) Temperature variation of the large, slow
 oscillation extracted by fitting Eq.~(\ref{eq:DHO}) to the
 temporal profiles. }\label{fig1}
\end{figure*}

SrTiO$_3$ is a popular material that is widely used for research
and in industries as a diamond simulant, substrates for
perovskytee-crystal growth, etc. It is well-known that SrTiO$_3$,
an insulator without doping, becomes metallic by slight doping of
Nb ions, exhibiting superconductor behaviour\cite{Koonce1967},
Also, phonon-related interesting phenomena in SrTiO$_3$ including
Poiseuille flow of hydrodynamic phonons\cite{Martelli2018} and
phonon hall effect\cite{Li2020} are suggested recently.
SrTiO$_3$ is also known as a quantum paraelectric that does not
undergo ferroelectric phase transition due to the relatively
large quantum fluctuations of a polar lattice vibration, the
so-called ferroelectric`soft mode'\cite{Burrett}. This is
advantageous for the existence of second sound because the
low-frequency, long-wavelength soft mode is thermally excited
even at cryogenic temperatures\cite{Gurevich-Tagantsev}.

The existence of second sound in SrTiO$_3$ was first
experimentally suggested by Hehlen et al\cite{Hehlen} according
to the anomalous Brillouin light scattering spectrum.  After a
12-year debate\cite{Scott-TPDS, Courtens-to-Scott,
Yamaguchi-prb-2002, Scott-myth}, Koreeda et al
\cite{Koreeda-PRL-2007, Koreeda-PRB-2009} reported that second
sound is likely to exist in SrTiO$_3$ from the analysis of the
Brillouin spectra based on extended irreversible
thermodynamics\cite{EIT-book}. However, as the lineshape analysis
was rather complicated, the existence of second sound in
SrTiO$_3$ has still been considered very
unlikely\cite{Scott-myth, Scott2013}, 
controversial\cite{Huberman2019}, or limited to an indirect
demonstration\cite{Ghosh2022a}.
For facilitating future applications of three-dimensional novel
thermal devices such as diodes, waveguides of heat or thermal
metamaterials, a convincing demonstration of SrTiO$_3$ as a 
medium of propagating second sound has long been awaited.

Here, we demonstrate the optical generation of
a coherent wave of a temperature field in SrTiO$_3$. This is the
first `direct' demonstration of  second-sound propagation in
SrTiO$_3$.
%
By systematically changing the wavelength and frequency of the
excitation generated in our transient-grating
method\cite{Maznev-OL-1998}, we find that the generated wave
excitation has a predicted velocity of second sound in
SrTiO$_3$.
We show that the generated wave excitation is overdamped at long
wavelengths and low frequencies, while at high temperatures, it
exhibits an apparent diffusive behaviour, with a predicted decay
rate of thermal diffusion in this material.
Combining our data with those reported in low-frequency light
scattering experiments at much shorter
wavelengths\cite{Tsujimi-doublet, Koreeda-PRL-2007,
Koreeda-PRB-2009}, we show that both the normal
(momentum-conserving) and resistive (momentum-destroying)
phonon-scattering rates can be easily obtained from  a
simple analysis of the wavenumber dependencies of the frequency
and damping rate of second sound. The obtained phonon-scattering
rates are in excellent agreement with previously reported
values\cite{Koreeda-PRL-2007, Koreeda-PRB-2009}, firmly verifying
that the `second-sound window' is widely open, i.e., the normal
scattering is much more frequent than the resistive scattering in 
cryogenic SrTiO$_3$.

\section*{FROM THERMAL DIFFUSION TO WAVE PROPAGATION}\label{sec2}

SrTiO$_3$ samples were synthesised single crystals. We used two
samples: Sample A and B. Sample A, which was very slightly amber
coloured, was purchased in 1995 and was identical to that
employed in previous studies\cite{Koreeda2006a, Koreeda-PRL-2007,
Koreeda-PRB-2009}. Sample B was purchased in 2022 and was
non-coloured and transparent. Both samples were blocks having
dimensions of $10\times8\times7$~mm$^3$, with surfaces oriented
to [001], [110], and [1$\bar 1$0] in the cubic phase.  The
transient grating was formed along the crystalline [001]
direction in the cubic phase, i.e., the wavevector $\vec{q}$ was
set parallel to $[001]_c$. The other direction of $\vec{q}$
eventually provided stronger damping, probably owing to the
anisotropy of the thermal conductivity in the tetragonal phase of
SrTiO$_3$.

\begin{figure*}[t]
 \centering \includegraphics[width=\textwidth]{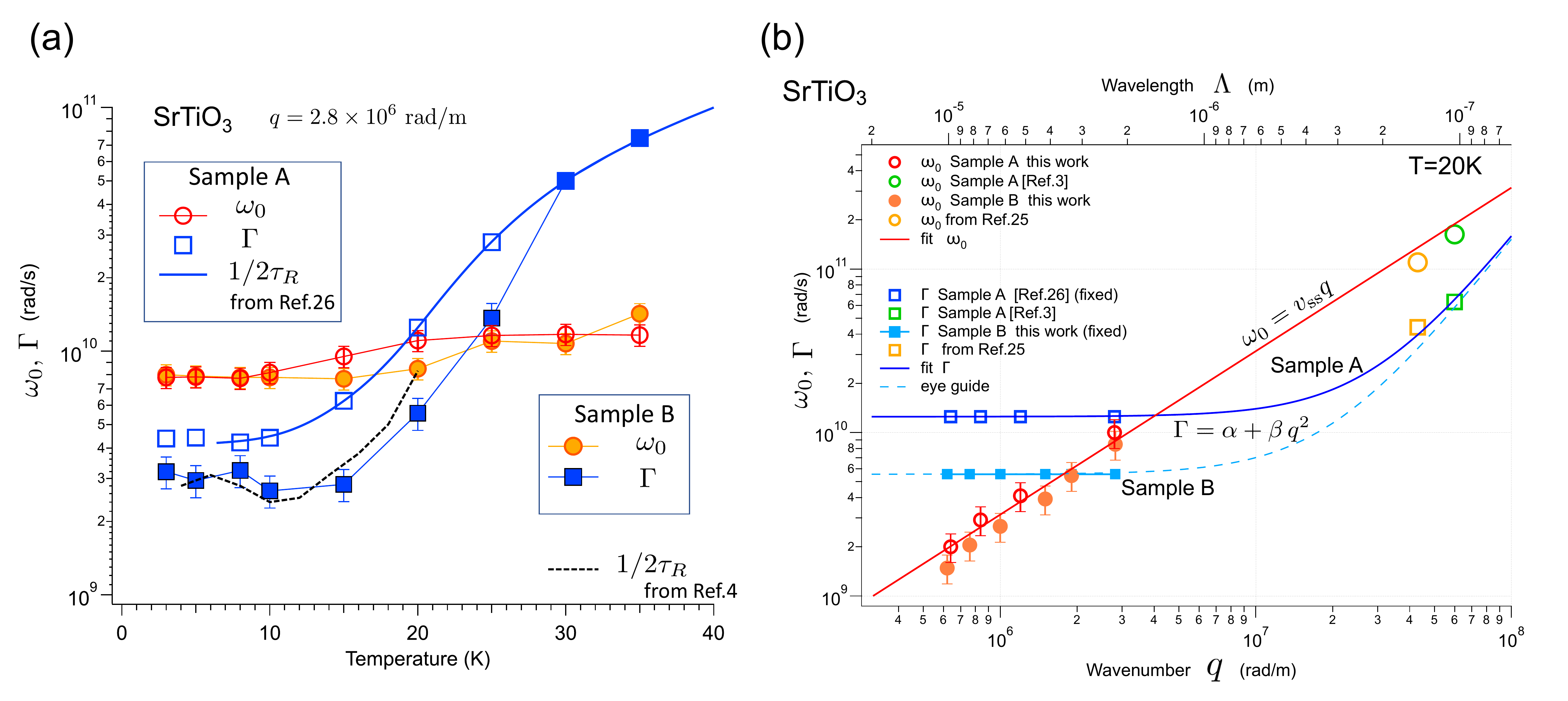}
 \caption{(a) Temperature dependences of the natural angular
 frequency ($\omega_0$) and damping rate ($\Gamma$) in Samples A
 and B. The literature values of $1/2\tau_R$ cited from
 Refs.\onlinecite{Koreeda2006a} and \onlinecite{Martelli2018} are
 also displayed. (b) Wavevector dependences of $\omega_0$ and
 $\Gamma$ in Samples A and B (log-log plot). The points at large
 $q-$values are cited from Refs.\onlinecite{Koreeda-PRL-2007} and
 \onlinecite{Tsujimi-doublet}. The red straight line and the blue
 curve are a fit to $\omega_0-$ and $\Gamma-$values obtained in
 Sample A, respectively. The dashed curve is a guide to the eye
 for $\Gamma$ in Sample B: it is drawn using $\alpha=5.6\times
 10^9$~radg/s and the same value of $\beta$ obtained for Sample
 A. }\label{fig:T+q-dep}
\end{figure*}


Figure \ref{fig1}a shows the temperature variation of the
temporal profiles, which were obtained through the
optically-heterodyne-detected thermal grating
technique\cite{Maznev-OL-1998} with a grating wavenumber of
2.8$\times 10^6$~rad/m (a grating period of 2.2$\times
10^{-6}$~m), after the subtraction of backgrounds (see Extended
Data Figs 1, 2, 3). The weak oscillation appearing on all the
traces is due to the longitudinal acoustic (LA) phonon mode,
which was identified from the known LA-sound velocity
($v_\text{LA}\approx 8.0 \times 10^3$~m/s) and employed
wavenumber $q$ according to the following dispersion relation:
$f= v_\text{LA}q/2\pi=3.6\times10^9$~Hz.

Our subject is the much more dominant component in the observed
traces.  The traces at relatively high temperatures in
Fig.~\ref{fig1}a appear to be exponential decays, which are
identified to be the thermal-diffusion signal as previously
reported\cite{Koreeda2006a}. We could overlay
$\exp(-D_\text{th}q^2 t)$ on the 35~K data with the reported
value of the thermal diffusivity $D_\text{th}=6.0\times
10^{-5}$~m$^2$/s for both Samples A and B, indicated by
the blue dashed curve on the 35~K trace in Fig.~\ref{fig1}a.
At relatively low temperatures, we see that the thermal-diffusive
signal transforms into underdamped oscillations with a period of
approximately 0.8~ns.
Figure \ref{fig1}b displays the spectra calculated by fast
Fourier transform (FFT) of the temporal profiles shown in
Fig.~\ref{fig1}a. The large oscillatory component yields a
resonance peak at approximately 1~GHz, whereas it vanishes at
high temperatures above 20~K.  The weak peak at approximately
3.8~GHz corresponds to the LA-phonon resonance, which has an
asymmetric ``Fano-type'' lineshape similar to that observed in
spontaneous Brillouin scattering\cite{Hehlen, Koreeda2006a,
Koreeda-PRL-2007}.

Each temporal profile was successfully fitted using a damped
harmonic oscillator (DHO) function as 
\begin{align}
 S(t)&\propto \frac{\omega_0}{\Omega}\,e^{-\Gamma t}
 \sin \Omega\, t, \quad \Omega\equiv \sqrt{\omega_0^2-\Gamma^2},
   \label{eq:DHO}
\end{align}
in addition to a weak LA-phonon oscillation (coherent LA phonon,
for details of analysis, see Methods and Extended Data Fig.~3).
Here, $\omega_0$ and $\Gamma$ are the natural angular frequency
and damping rate of the DHO, respectively.  The fitted functions
are overlaid on the experimental points in Fig.~\ref{fig1}a as
black curves. Figure~\ref{fig1}c displays the extracted response
corresponding to the dominant component expressed by
Eq.~(\ref{eq:DHO}). As we have already identified that the
high-temperature traces are due to thermal diffusion, which is
the `overdamped thermal wave', the generation of oscillating
signals must be attributed to underdamped wave propagation of
heat, i.e., second sound.

\section*{DISCUSSION}
 Figure~\ref{fig:T+q-dep}a indicates the temperature dependences
of the natural angular frequency $\omega_0$ obtained by function
fitting the function in Eq.~(\ref{eq:DHO}) to the observed
temporal profiles as well as the values of $\Gamma$ employed in
our analysis.
The fact that thermal diffusion is observed in our experiments
allows us to assume that the hydrodynamic-phonon
regime\cite{Koreeda-PRB-2009} is realised in our experimental
condition.  Thus, we may assume the damping rate of the thermal
wave for small wavenumbers to be $\Gamma\approx
1/2\tau_R=\bar{v}_\text{ph}^2/6D_\text{th}$\cite{Koreeda-PRB-2009},
where $\tau_R$ is the resistive phonon scattering time, and
$\bar{v}_\text{ph}$ is the averaged phonon velocity.  As Sample A
is exactly the same as that employed in
Ref.\onlinecite{Koreeda2006a}, we can use the (interpolated)
value of $D_\text{th}$ reported in that study 
at all the temperatures investigated in the present study, except
3~K and 5~K.  With the $\Gamma$-value fixed in the DHO-function
fitting at each temperature, the values of $\omega_0$ were
successfully determined with minimum uncertainty in Sample A.
For Sample B, we could employ the same $\Gamma$-values as in
Sample A only for 30~K and 35~K. For lower temperatures, however,
Sample B exhibited more weakly damped oscillations than Sample A,
and we could find appropriate values for $\Gamma$, by assuming a
weak and smooth temperature dependence of $\omega_0$, as in
Sample A. The underdamped signals (decaying oscillations) can be
successfully fitted with both $\omega_0$ and $\Gamma$ being
independent free parameters, because the temporal shape contains
independent information of $\omega_0$ and $\Gamma$, which are
relevant for obtaining the oscillation period and decaying rate,
respectively. We also plotted the values of $1/2\tau_R$
calculated from the values of mean free path reported by Martelli
et al\cite{Martelli2018} in Fig.~\ref{fig:T+q-dep}a, which are in
good agreement with $\Gamma$ obtained in Sample B at $T\lesssim
20$~K.

Figure~\ref{fig:T+q-dep}a shows that $T \approx 18~$K and
$T\approx 23~$K are the critical temperatures between overdamping
(thermal diffusion) and underdamping (thermal wave) in Samples A
and B, respectively, for the employed value of $q$.
We obtained relatively small values of $\Gamma$ in Sample B
compared to those in Sample A. This is probably due to the worse
thermal conductivity resulting from the sample
quality\cite{Martelli2018} with more concentration of the oxygen
vacancies or impurities, which is apparent from the colour of
Sample A (see Sample). Conversely, we obtained similar values of
$\omega_0$ in both samples. This is because $\omega_0$ is
basically associated with the average phonon velocity or the
speed of second sound (described below), which is determined
solely by the phonon dispersion relation (common to both samples)
rather than the scattering rate of phonons.

If the generated oscillation is due to a wave, the frequency
$\omega_0$ and wavenumber $q$ must obey the dispersion relation
as $\omega_0 = q v$. where $v$ is the phase velocity of the
wave. It is known that second sound has a velocity given by
$v_\text{ss}\approx \frac1{\sqrt{3}} \bar v_\text{ph}$.  With the
known value of $\bar{v}_\text{ph}\approx 5.4\times 10^3$~m/s
(from Debye's model) for SrTiO$_3$, we estimate that
$v_\text{ss}\approx 3.1\times 10^3$~m/s. We also investigated the
wavenumber ($q$) dependence of the generated excitation (see
Extended data Fig.4).
Since we investigated the $q-$dependence at a fixed temperature,
viz., $T=20$~K, $\Gamma$ was fixed when analysing
$q-$dependence. We set the $\Gamma$ value to be $1.3\times
10^9$~rad/s and $5.6\times 10^9$~rad/s for Samples A and
B, respectively (Fig.~\ref{fig:T+q-dep}a).  Regardless of
employing this {\em $q$-independent} $\Gamma$, each of the
temporal profiles observed at various $q-$values was successfully
fitted again using the DHO function of Eq.~(\ref{eq:DHO}) plus
the weak LA-phonon oscillation (Extended data Fig.4).

Figure \ref{fig:T+q-dep}b shows the $q-$dependence of $\omega_0$
measured at 20~K in Samples A and B, as well as the values
reported in the spectral analyses for the broad
Brillouin-scattered peak measured at $\approx$20~K, presented in
Refs.~\onlinecite{Koreeda-PRL-2007, Tsujimi-JKCFE06}.
We clearly see that $\omega_0\propto q$, which is  a linear
dispersion relation, indicating that the oscillation arises from
a wave excitation of wavelengths $\Lambda=2\pi/q$, and the slope
of the plots provides the propagation velocity of the
wave. Furthermore, the points cited from
Refs.~\onlinecite{Koreeda-PRL-2007} and
\onlinecite{Tsujimi-JKCFE06} are located on a common line with
the points obtained in this work, indicating that the generated
wave-excitation has the same physical origin as the broad
Brillouin-scattered peak.
From a line fitting to all of the $\omega_0$-values against $q$
for Sample A including the data for much larger wavenumbers
($q=4.2\times 10^7$ [Ref.\onlinecite{Tsujimi-JKCFE06}], and
$6.0\times 10^7$ rad/m [Ref.\onlinecite{ Koreeda-PRL-2007}]), the
velocity of the wave was found to be $\approx 3.1\times 10^3$
m/s, which is in excellent agreement with the predicted
second-sound velocity.
We also plot the values of the damping rates $\Gamma$ in
Fig.~\ref{fig:T+q-dep}b. Because $\Gamma$ is constant at small
$q$ values, the magnitudes of $\omega_0$ and $\Gamma$ cross at
approximately a `critical wavenumber' of $q_c\approx 4.0 \times 10^6$
and $2.0\times 10^6$ rad/m for Samples A and B, respectively,
i.e., $\omega_0>\Gamma$ for $q>q_c$, and vice versa. 
The generated signal became `diffusive' when $q\ll q_c$,
and oscillatory when $q>q_c$ (see Extended Data Fig.4). This
behaviour simply indicates that the generated thermal
wave can exist as a weakly damped (underdamped) wave if
$\omega_0>\Gamma$ is satisfied.

For $\Gamma$ at larger values of $q$, we again cite the published
data based on the spectral analyses performed in studies on
spontaneous Brillouin scattering\cite{Koreeda-PRL-2007,
Tsujimi-JKCFE06}. For the damping rate $\Gamma$ in Sample A, the
$q$ dependence is reasonably fitted by the predicted
relation for second sound as\cite{Koreeda-PRB-2009}
\begin{align}
 \Gamma= \alpha+ \beta\,q^2,\quad
 \alpha=\frac1{2\tau_R},\,\,\beta=\dfrac2{5}v_\text{ss}^2\tau_N
 \label{eq:Gamma}
\end{align}
where $\tau_R$ and $\tau_N$ are the resistive and normal
scattering times of the phonons, respectively.
%
%
With a fixed value of $\alpha$, fitting of  a  parabola
as simple as Eq.~(\ref{eq:Gamma}) to the $q$ dependence of $\Gamma$
directly gives us the values of $\tau_N$, with the knowledge
of the second-sound velocity $v_\text{ss} = 3.1\times 10^3$ m/s,
which was just obtained from the dispersion relation,
$\omega_0=v_\text{ss}q$.
For Sample A, the constant $\alpha$ was fixed to $1.3\times
10^{10}$~rad/s according to the separately measured value of
$\tau_R=3D_\text{th}/\bar{v}_\text{ph}^2= 40$~ps for Sample A
\cite{Koreeda2006a}. We could fit the blue curve in
Fig.~\ref{fig:T+q-dep}b to obtain $\beta=1.5\times
10^{-5}$~m$^2$/s$^2$rad, yielding $\tau_N= 3.9$~ps, which is in
fair agreement with the previously reported value ($\approx
5$~ps) for SrTiO$_3$ at 20~K\cite{Koreeda-PRL-2007,
Koreeda-PRB-2009}, indicating that the frequency window required
for the existence of second sound\cite{Ashcroft-Mermin} is open
in Sample A.
For Sample B, we may assume that $\tau_N$ adopts a similar value
as Sample A, because unlike $\tau_R$, $\tau_N$ should not be
affected by the sample quality and can be determined essentially
by the frequency of the transverse optic phonon mode (soft
mode)\cite{Gurevich-Tagantsev}.  We plot in
Fig.~\ref{fig:T+q-dep}b an eye-guide curve (dashed curve)
employing the same $\beta$ value as in Sample A. Because the
magnitude of $\Gamma$ in Sample B is approximately half that of
$\Gamma$ in Sample A, we estimate $\tau_R\approx 90$~ps. Thus,
the frequency window for second-sound propagation is even more
widely open in newer crystals of SrTiO$_3$ such as Sample B.

In terms of length scale, the mean free paths of normal and
resistive phonons ($\bar{v}_\text{ph}\tau_N$ and
$\bar{v}_\text{ph}\tau_R$, respectively) in Sample B at 20~K are
estimated to be 21~nm and 490~nm, respectively, indicating that
second sound having a wavelength between these values can
propagate as a wave. As the reported size of the
anti-ferrodistortive domains arising from the cubic-to-tetragonal
structural phase transition at 105~K in SrTiO$_3$ is in the $\mu$m
scale\cite{Buckley1999}, the second sound propagating along
$[001]_c$ direction in cryogenic SrTiO$_3$ does not suffer much
from domain wall scattering of phonons.
Here, it is worth noting that we have directly obtained the value
of $\tau_N$ just by fitting of quite a simple parabola function,
instead of employing a complicated fitting function like the ones
presented in previous studies\cite{Koreeda-PRL-2007,
Koreeda-PRB-2009}.

The $q-$dependence of the damping rate $\Gamma$ presented above
is a prominent feature and direct evidence of second
sound. Although acoustic phonons are the only alternative
candidate for the generated wave in this frequency range, they
must exhibit a damping rate of $\Gamma_\text{ph} \propto q^p$,
where the exponent $p$ approaches 2 for small $q$ and high $T$
(Akhieser limit), and approaches 1 for large $q$ and low $T$
(Landau-Rumer limit)\cite{Ohno2017}. Therefore, the observed
$q$ independence for small $q-$values and steeper $q-$dependence
for large $q-$values is not compatible with the scenario of the
attenuation of an acoustic-phonon mode\cite{Nakamura2015}.

The temperature and wavevector dependences shown in
Figs.~\ref{fig:T+q-dep}a and b unambiguously confirms that
the generated wave excitation is the `coherent second sound' in
SrTiO$_3$. Thus, we have succeeded in coherently generating a
temperature wave with well-defined and controllable frequencies and
phases.
Simultaneously, the 27 year-controversy\cite{Hehlen, Scott-TPDS,
Courtens-to-Scott, Yamaguchi-prb-2002, Scott-myth,
Koreeda-PRL-2007, Koreeda-PRB-2009, Scott2013, Nakamura2015,
Huberman2019} over the broad Brillouin-scattered peak has been
resolved, and it has been firmly identified as the spectrum of
the `thermally excited' second sound in SrTiO$_3$.  The FFT
spectra shown in Fig.~\ref{fig1}b and Extended data Fig.~4
qualitatively exhibit the same lineshape as that reported in the
spontaneous Brillouin scattering experiments\cite{Hehlen,
Koreeda-PRL-2007, Koreeda-PRB-2009}.

\section{SUMMARY AND OUTLOOK}

We have succeeded in generating a coherent heat wave (second
sound) in SrTiO$_3$ using the stimulated Brillouin scattering
technique.
We emphasise the difference between observing thermally-excited
second sound and generating coherent second sound. The broad
Brillouin peak arises from the thermally-excited, `incoherent'
ripples of temperature fluctuations with extremely small
amplitudes, whereas a laser-pulse-induced stimulated scattering
process actively generates `coherent' (in-phase) oscillations
of temperature with much larger amplitudes. Our experimental
achievement may open a way for controlling such eigenstates of
heat. For example, we may be able to establish various initial
patterns of temperature rise by changing the figures displayed on
the spatial light modulator employed in our apparatus (see
Methods and Extended Data Fig.1).

Although no experimental data on the $q$ value range from
$1\times 10^7$ to $\sim3 \times 10^7$ rad/m have reported thus
far, we find that the ratio $\omega_0/\Gamma$ is the largest for
this $q$ range.  In fact, $\omega_0/\Gamma$ is expected to reach
its maximum of $\approx 6$ at $q=\sqrt{\alpha/\beta} \approx
1.8\times 10^7$ rad/m, which corresponds to a second-sound
wavelength of $\Lambda = 2\pi/q\approx 350$~nm. At this
wavelength, the second-sound resonance in SrTiO$_3$ is expected
to be the sharpest.  Such a well-defined wave excitation would be
sufficient for various kinds of wave-control approaches of heat.
Because SrTiO$_3$ is a 3D bulk crystal, wave-controlling elements
such as reflecting interfaces (mirrors), lenses, or waveguides,
seem to be easily constructed with SrTiO$_3$.  Fortunately, since
SrTiO$_3$ is a very chemically-stable oxide and an industrially
ubiquitous material, high-quality single crystals are widely
available in market.  Also, SrTiO$_3$ thin films with thicknesses
of the second-sound wavelength, namely, 100$\sim$500~nm, may be
fabricated on various substrate materials.
Superconductivity of doped SrTiO$_3$\cite{Gastiasoro2020} would
be more interesting if the superconducting phase transition could
be externally modulated by coherent temperature waves.
We expect many attractive applications of thermal wave utilising
SrTiO$_3$. 




\bibliographystyle{naturemag}
\bibliography{../../../../mylibrary}

\begin{thebibliography}{10}
\expandafter\ifx\csname url\endcsname\relax
  \def\url#1{\texttt{#1}}\fi
\expandafter\ifx\csname urlprefix\endcsname\relax\def\urlprefix{URL }\fi
\providecommand{\bibinfo}[2]{#2}
\providecommand{\eprint}[2][]{\url{#2}}

\bibitem{Gurevich-Tagantsev}
\bibinfo{author}{Gurevich, V.~L.} \& \bibinfo{author}{Tagantsev, A.~K.}
\newblock \bibinfo{title}{{Second Sound in Ferroelectircs}}.
\newblock \emph{\bibinfo{journal}{Sov. Phys. JETP}}
  \textbf{\bibinfo{volume}{67}}, \bibinfo{pages}{206--212}
  (\bibinfo{year}{1988}).

\bibitem{Hehlen}
\bibinfo{author}{Hehlen, B.}, \bibinfo{author}{P{\'{e}}rou, A.-L.},
  \bibinfo{author}{Courtens, E.} \& \bibinfo{author}{Vacher, R.}
\newblock \bibinfo{title}{{Observation of a Doublet in the Quasielastic Central
  Peak of Quantum-Paraelectric SrTi$O_{3}$}}.
\newblock \emph{\bibinfo{journal}{Phys. Rev. Lett.}}
  \textbf{\bibinfo{volume}{75}}, \bibinfo{pages}{2416--2419}
  (\bibinfo{year}{1995}).

\bibitem{Koreeda-PRL-2007}
\bibinfo{author}{Koreeda, A.}, \bibinfo{author}{Takano, R.} \&
  \bibinfo{author}{Saikan, S.}
\newblock \bibinfo{title}{{Second Sound in {SrTiO$_3$}}}.
\newblock \emph{\bibinfo{journal}{Physical Review Letters}}
  \textbf{\bibinfo{volume}{99}}, \bibinfo{pages}{265502(1--4)}
  (\bibinfo{year}{2007}).

\bibitem{Martelli2018}
\bibinfo{author}{Martelli, V.}, \bibinfo{author}{Jim{\'{e}}nez, J.~L.},
  \bibinfo{author}{Continentino, M.}, \bibinfo{author}{Baggio-Saitovitch, E.}
  \& \bibinfo{author}{Behnia, K.}
\newblock \bibinfo{title}{{Thermal Transport and Phonon Hydrodynamics in
  Strontium Titanate}}.
\newblock \emph{\bibinfo{journal}{Physical Review Letters}}
  \textbf{\bibinfo{volume}{120}}, \bibinfo{pages}{1--6} (\bibinfo{year}{2018}).
\newblock
  \urlprefix\url{https://link.aps.org/doi/10.1103/PhysRevLett.120.125901}.

\bibitem{Huberman2019}
\bibinfo{author}{Huberman, S.} \emph{et~al.}
\newblock \bibinfo{title}{{Observation of second sound in graphite at
  temperatures above 100 K}}.
\newblock \emph{\bibinfo{journal}{Science (New York, N.Y.)}}
  \textbf{\bibinfo{volume}{3548}}, \bibinfo{pages}{1--15}
  (\bibinfo{year}{2019}).
\newblock \urlprefix\url{http://arxiv.org/abs/1901.09160
  http://www.ncbi.nlm.nih.gov/pubmed/30872535}.
\newblock \eprint{1901.09160}.

\bibitem{Li2020}
\bibinfo{author}{Li, X.}, \bibinfo{author}{Fauqu\'e, B.}, \bibinfo{author}{Zhu,
  Z.} \& \bibinfo{author}{Behnia, K.}
\newblock \bibinfo{title}{Phonon thermal hall effect in strontium titanate}.
\newblock \emph{\bibinfo{journal}{Phys. Rev. Lett.}}
  \textbf{\bibinfo{volume}{124}}, \bibinfo{pages}{105901}
  (\bibinfo{year}{2020}).
\newblock
  \urlprefix\url{https://link.aps.org/doi/10.1103/PhysRevLett.124.105901}.

\bibitem{Ishito2021}
\bibinfo{author}{Ishito, K.} \emph{et~al.}
\newblock \bibinfo{title}{Truly chiral phonons in $\alpha$-hgs}
  (\bibinfo{year}{2021}).
\newblock \urlprefix\url{https://arxiv.org/abs/2110.11604}.

\bibitem{Oishi2022}
\bibinfo{author}{Oishi, E.}, \bibinfo{author}{Fujii, Y.} \&
  \bibinfo{author}{Koreeda, A.}
\newblock \bibinfo{title}{Selective observation of enantiomeric chiral phonons
  in $\alpha$-quartz} (\bibinfo{year}{2022}).
\newblock \urlprefix\url{https://arxiv.org/abs/2210.07526}.

\bibitem{Ghosh2022a}
\bibinfo{author}{Ghosh, K.}, \bibinfo{author}{Kusiak, A.} \&
  \bibinfo{author}{Battaglia, J.-L.}
\newblock \bibinfo{title}{{Phonon hydrodynamics in crystalline materials}}.
\newblock \emph{\bibinfo{journal}{Journal of Physics: Condensed Matter}}
  \textbf{\bibinfo{volume}{34}}, \bibinfo{pages}{323001}
  (\bibinfo{year}{2022}).
\newblock \eprint{2205.11345}.

\bibitem{Ackerman-PRL-1969}
\bibinfo{author}{Ackerman, C.~C.} \& \bibinfo{author}{Overton, W.~C.}
\newblock \bibinfo{title}{{Second Sound in Solid Helium-3}}.
\newblock \emph{\bibinfo{journal}{Phys. Rev. Lett.}}
  \textbf{\bibinfo{volume}{22}}, \bibinfo{pages}{764--766}
  (\bibinfo{year}{1969}).

\bibitem{SS-in-Bi}
\bibinfo{author}{Narayanamurti, V.} \& \bibinfo{author}{Dynes, R.~C.}
\newblock \bibinfo{title}{{Observation of Second Sound in Bismuth}}.
\newblock \emph{\bibinfo{journal}{Phys. Rev. Lett.}}
  \textbf{\bibinfo{volume}{28}}, \bibinfo{pages}{1461--1465}
  (\bibinfo{year}{1972}).

\bibitem{Jackson1970}
\bibinfo{author}{Jackson, H.~E.}, \bibinfo{author}{Walker, C.~T.} \&
  \bibinfo{author}{McNelly, T.~F.}
\newblock \bibinfo{title}{{Second Sound in {NaF}}}.
\newblock \emph{\bibinfo{journal}{Phys. Rev. Lett.}}
  \textbf{\bibinfo{volume}{25}}, \bibinfo{pages}{26--28}
  (\bibinfo{year}{1970}).

\bibitem{Ding2022a}
\bibinfo{author}{Ding, Z.} \emph{et~al.}
\newblock \bibinfo{title}{{Observation of second sound in graphite over 200
  K}}.
\newblock \emph{\bibinfo{journal}{Nature Communications}}
  \textbf{\bibinfo{volume}{13}}, \bibinfo{pages}{1--9} (\bibinfo{year}{2022}).

\bibitem{Scott-TPDS}
\bibinfo{author}{Scott, J.~F.} \& \bibinfo{author}{Ledbetter, H.}
\newblock \bibinfo{title}{{Interpretation of elastic anomalies in {SrTiO$_3$}
  at 37K}}.
\newblock \emph{\bibinfo{journal}{Z. Phys.}} \textbf{\bibinfo{volume}{B 104}},
  \bibinfo{pages}{635--639} (\bibinfo{year}{1997}).
\newblock
  \urlprefix\url{http://www.springerlink.com/content/u8wuvcl8f5encg1f/}.

\bibitem{Courtens-to-Scott}
\bibinfo{author}{Courtens, E.}, \bibinfo{author}{Hehlen, B.},
  \bibinfo{author}{Farhi, E.} \& \bibinfo{author}{Tagantsev, A.~K.}
\newblock \bibinfo{title}{{Optical mode crossings and the low temperature
  anomalies of {SrTiO$_3$}}}.
\newblock \emph{\bibinfo{journal}{Zeitschrift f{\"{u}}r Physik B Condensed
  Matter}} \textbf{\bibinfo{volume}{104}}, \bibinfo{pages}{641--642}
  (\bibinfo{year}{1997}).
\newblock
  \urlprefix\url{http://192.129.24.144/licensed_materials/00257/bibs/7104004/71040641.htm}.

\bibitem{Yamaguchi-prb-2002}
\bibinfo{author}{Yamaguchi, M.} \emph{et~al.}
\newblock \bibinfo{title}{{Brillouin-scattering study of the broad doublet in
  isotopically exchanged {SrTiO$_{3}$}}}.
\newblock \emph{\bibinfo{journal}{Phys. Rev. B}} \textbf{\bibinfo{volume}{65}},
  \bibinfo{pages}{172102} (\bibinfo{year}{2002}).

\bibitem{Scott-myth}
\bibinfo{author}{Scott, J.~F.}, \bibinfo{author}{Chen, A.} \&
  \bibinfo{author}{Ledbetter, H.}
\newblock \bibinfo{title}{{The myth of second sound in strontium titanate}}.
\newblock \emph{\bibinfo{journal}{Journal of Physics and Chemistry of Solids}}
  \textbf{\bibinfo{volume}{61}}, \bibinfo{pages}{185--190}
  (\bibinfo{year}{2000}).
\newblock
  \urlprefix\url{http://www.sciencedirect.com/science/article/B6TXR-3Y0J888-8/2/445c9018ed83098ed739d5d8d5af2921}.

\bibitem{Koreeda-PRB-2009}
\bibinfo{author}{Koreeda, A.}, \bibinfo{author}{Takano, R.} \&
  \bibinfo{author}{Saikan, S.}
\newblock \bibinfo{title}{{Light scattering in a phonon gas}}.
\newblock \emph{\bibinfo{journal}{Physical Review B}}
  \textbf{\bibinfo{volume}{80}}, \bibinfo{pages}{165104(1--25)}
  (\bibinfo{year}{2009}).
\newblock \urlprefix\url{https://link.aps.org/doi/10.1103/PhysRevB.80.165104}.

\bibitem{Scott2013}
\bibinfo{author}{Scott, J.~F.}
\newblock \bibinfo{title}{Prospects for ferroelectrics: 2012-2022}.
\newblock \emph{\bibinfo{journal}{ISRN Materials Science}}
  \textbf{\bibinfo{volume}{2013}}, \bibinfo{pages}{187313}
  (\bibinfo{year}{2013}).
\newblock \urlprefix\url{https://doi.org/10.1155/2013/187313}.

\bibitem{Nakamura2015}
\bibinfo{author}{Nakamura, S.} \& \bibinfo{author}{Tsujimi, Y.}
\newblock \bibinfo{title}{Broad doublet spectra in the quantum paraelectric
  state of srtio3}.
\newblock \emph{\bibinfo{journal}{Ferroelectrics}}
  \textbf{\bibinfo{volume}{485}}, \bibinfo{pages}{20--26}
  (\bibinfo{year}{2015}).
\newblock \urlprefix\url{https://doi.org/10.1080/00150193.2015.1060086}.
\newblock \eprint{https://doi.org/10.1080/00150193.2015.1060086}.

\bibitem{Koonce1967}
\bibinfo{author}{Koonce, C.~S.}, \bibinfo{author}{Cohen, M.~L.},
  \bibinfo{author}{Schooley, J.~F.}, \bibinfo{author}{Hosler, W.~R.} \&
  \bibinfo{author}{Pfeiffer, E.~R.}
\newblock \bibinfo{title}{Superconducting transition temperatures of
  semiconducting srti${\mathrm{o}}_{3}$}.
\newblock \emph{\bibinfo{journal}{Phys. Rev.}} \textbf{\bibinfo{volume}{163}},
  \bibinfo{pages}{380--390} (\bibinfo{year}{1967}).
\newblock \urlprefix\url{https://link.aps.org/doi/10.1103/PhysRev.163.380}.

\bibitem{Burrett}
\bibinfo{author}{Barrett, J.~H.}
\newblock \bibinfo{title}{{Dielectric Constant in Perovskite Type Crystals}}.
\newblock \emph{\bibinfo{journal}{Phys. Rev.}} \textbf{\bibinfo{volume}{86}},
  \bibinfo{pages}{118--120} (\bibinfo{year}{1952}).

\bibitem{EIT-book}
\bibinfo{author}{Jou, D.}, \bibinfo{author}{Casas-V{\'{a}}zquez, J.} \&
  \bibinfo{author}{Lebon, G.}
\newblock \emph{\bibinfo{title}{{Extended Irreversible Thermodynamics}}}
  (\bibinfo{publisher}{Springer}, \bibinfo{address}{Berlin},
  \bibinfo{year}{2001}), \bibinfo{edition}{3rd rev ed} edn.

\bibitem{Maznev-OL-1998}
\bibinfo{author}{Maznev, A.~A.}, \bibinfo{author}{Nelson, K.~A.} \&
  \bibinfo{author}{Rogers, J.~A.}
\newblock \bibinfo{title}{{Optical heterodyne detection of laser-induced
  gratings}}.
\newblock \emph{\bibinfo{journal}{Opt. Lett.}} \textbf{\bibinfo{volume}{23}},
  \bibinfo{pages}{1319--1321} (\bibinfo{year}{1998}).
\newblock \urlprefix\url{http://ol.osa.org/abstract.cfm?URI=ol-23-16-1319
  http://www.ncbi.nlm.nih.gov/pubmed/18087511}.

\bibitem{Tsujimi-doublet}
\bibinfo{author}{Tan, M.}, \bibinfo{author}{Tsujimi, Y.} \&
  \bibinfo{author}{Yagi, T.}
\newblock \bibinfo{title}{{Broad Doublet Spectra in {SrTiO$_3$}}}.
\newblock \emph{\bibinfo{journal}{Journal of the Korean Physical Society}}
  \textbf{\bibinfo{volume}{46}}, \bibinfo{pages}{97--99}
  (\bibinfo{year}{2005}).

\bibitem{Koreeda2006a}
\bibinfo{author}{Koreeda, A.}, \bibinfo{author}{Nagano, T.},
  \bibinfo{author}{Ohno, S.} \& \bibinfo{author}{Saikan, S.}
\newblock \bibinfo{title}{{Quasielastic light scattering in rutile, ZnSe ,
  silicon, and SrTiO$_3$}}.
\newblock \emph{\bibinfo{journal}{Physical Review B}}
  \textbf{\bibinfo{volume}{73}}, \bibinfo{pages}{024303}
  (\bibinfo{year}{2006}).
\newblock \urlprefix\url{https://link.aps.org/doi/10.1103/PhysRevB.73.024303}.

\bibitem{Tsujimi-JKCFE06}
\bibinfo{author}{Tsujimi, Y.} \& \bibinfo{author}{Itoh, M.}
\newblock \bibinfo{title}{{Broad Doublet Spectra Observed in Strontium
  Titanate}}.
\newblock \emph{\bibinfo{journal}{Journal of the Korean Physical Society}}
  \textbf{\bibinfo{volume}{51}}, \bibinfo{pages}{819--823}
  (\bibinfo{year}{2007}).

\bibitem{Ashcroft-Mermin}
\bibinfo{author}{Ashcroft, N.~W.} \& \bibinfo{author}{Mermin, N.}
\newblock \emph{\bibinfo{title}{{Solid State Physics}}}
  (\bibinfo{publisher}{Thomson Learning}, \bibinfo{address}{London},
  \bibinfo{year}{1976}).

\bibitem{Buckley1999}
\bibinfo{author}{Buckley, A.}, \bibinfo{author}{Rivera, J.~P.} \&
  \bibinfo{author}{Salje, E. K.~H.}
\newblock \bibinfo{title}{Twin structures in tetragonal srtio3: The
  ferroelastic phase transition and the formation of needle domains}.
\newblock \emph{\bibinfo{journal}{Journal of Applied Physics}}
  \textbf{\bibinfo{volume}{86}}, \bibinfo{pages}{1653--1656}
  (\bibinfo{year}{1999}).
\newblock \urlprefix\url{https://doi.org/10.1063/1.370942}.
\newblock \eprint{https://doi.org/10.1063/1.370942}.

\bibitem{Ohno2017}
\bibinfo{author}{Ohno, S.}, \bibinfo{author}{Sonehara, T.},
  \bibinfo{author}{Tatsu, E.}, \bibinfo{author}{Koreeda, A.} \&
  \bibinfo{author}{Saikan, S.}
\newblock \bibinfo{title}{{Attenuation process of the longitudinal phonon mode
  in a ${\mathrm{TeO}}_{2}$ crystal in the 20-GHz range}}.
\newblock \emph{\bibinfo{journal}{Phys. Rev. B}} \textbf{\bibinfo{volume}{95}},
  \bibinfo{pages}{224301} (\bibinfo{year}{2017}).
\newblock \urlprefix\url{https://link.aps.org/doi/10.1103/PhysRevB.95.224301}.

\bibitem{Gastiasoro2020}
\bibinfo{author}{Gastiasoro, M.~N.}, \bibinfo{author}{Ruhman, J.} \&
  \bibinfo{author}{Fernandes, R.~M.}
\newblock \bibinfo{title}{{Superconductivity in dilute SrTiO3: A review}}.
\newblock \emph{\bibinfo{journal}{Annals of Physics}}
  \textbf{\bibinfo{volume}{417}} (\bibinfo{year}{2020}).
\newblock \eprint{1912.01509}.

\bibitem{Enns1969}
\bibinfo{author}{Enns, R.~H.} \& \bibinfo{author}{Batra, I.~P.}
\newblock \bibinfo{title}{{Stimulated Thermal Scattering in the Second-Sound
  Regime}}.
\newblock \emph{\bibinfo{journal}{Phys. Rev.}} \textbf{\bibinfo{volume}{180}},
  \bibinfo{pages}{227--232} (\bibinfo{year}{1969}).

\bibitem{Batra1969}
\bibinfo{author}{Batra, I.~P.}, \bibinfo{author}{Enns, R.~H.},
  \bibinfo{author}{ENNS, I. P.~B.} \& \bibinfo{author}{H., R.}
\newblock \bibinfo{title}{{Stimulated thermal scattering of picosecond laser
  pulses in the second sound limit}}.
\newblock \emph{\bibinfo{journal}{Physics Letters A}}
  \textbf{\bibinfo{volume}{60}}, \bibinfo{pages}{63--64}
  (\bibinfo{year}{1969}).
\newblock
  \urlprefix\url{http://www.sciencedirect.com/science/article/B6TVM-46TY11K-1M/2/823ec8e73eea9972a17b06799362b26a}.

\bibitem{Muller-1979}
\bibinfo{author}{M{\"{u}}ller, K.~A.} \& \bibinfo{author}{Burkard, H.}
\newblock \bibinfo{title}{{{SrTiO$_{3}$}: An intrinsic quantum paraelectric
  below 4 K}}.
\newblock \emph{\bibinfo{journal}{Phys. Rev. B}} \textbf{\bibinfo{volume}{19}},
  \bibinfo{pages}{3593--3602} (\bibinfo{year}{1979}).

\bibitem{Koreeda2010}
\bibinfo{author}{Koreeda, A.}, \bibinfo{author}{Takano, R.},
  \bibinfo{author}{Ushio, A.} \& \bibinfo{author}{Saikan, S.}
\newblock \bibinfo{title}{{Collective phonon excitation in {KTaO$_3$}}}.
\newblock \emph{\bibinfo{journal}{Physical Review B}}
  \textbf{\bibinfo{volume}{82}}, \bibinfo{pages}{125103(1--7)}
  (\bibinfo{year}{2010}).

\end{thebibliography}




\newpage

\section{Methods}

\subsection*{Stimulated Brillouin scattering by second sound} 

We used transient grating method for thermal wave (second sound)
 in dielectric
solids. The measurement principle is described in
Refs.~\onlinecite{Enns1969} and \onlinecite{Batra1969}. As SrTiO$_3$ is an
incipient ferroelectric, the dielectric constant has a rather
strong temperature dependence\cite{Muller-1979}. Therefore,
thermal grating in SrTiO$_3$ is dominantly formed by coupling
of the refractive index or dielectric constant with 
temperature field, viz., $\therm{n}{T}{\rho}$ or
$\therm{\epsilon}{T}{\rho}$, 
rather than by coupling with density., viz., $\therm{n}{\rho}{T}$ or
$\therm{\epsilon}{\rho}{T}$.  The incident probe electric field
$E(\vec{r},t)$ is diffracted as a consequence of the nonlinear
polarization due to the temperature fluctuation as
\begin{align}
 P_\text{NL}(t) \propto \therm{\epsilon}{T}{\rho} \delta T (t)
 E(t), 
\end{align}
which leads to a gain factor $G(\omega)$ as\cite{Batra1969}
\begin{align}
 G(\omega) \propto \frac{\alpha_2 E^4}{\rho C_V }\therm{n}{T}{\rho}
 \frac{\omega \Gamma}{(\omega^2-\omega_0^2)^2+4\omega^2\Gamma^2},
 \label{eq: gain}
\end{align}
where $\alpha_2$ is the two-photon absorption coefficient, $C_V$
is the specific heat at constant volume, $\omega_0=v_\text{ss}q$
is the eigenfrequency of second sound without damping,
$v_\text{ss}$ is the second-sound velocity, and $\Gamma$ is the
damping rate of the second sound. Here, we considered
two-photon absorption\cite{Koreeda2006a} instead of linear
absorption, because SrTiO$_3$ is transparent for 532~nm radiation
of the pump laser employed in our experiments.
Accordingly, the corresponding time dependence of the diffracted
electric field can be calculated as Fourier transform of
Eq.~(\ref{eq: gain}) as 
\begin{align}
 P_\text{NL}(t) \propto 
 \frac{\alpha_2 E^4}{\rho C_V }\therm{n}{T}{\rho}
 \times e^{-\Gamma t}\,\, \frac{\sin \sqrt{\omega_0^2-\Gamma^2}\,\,t}
 {\sqrt{\omega_0^2-\Gamma^2}}.
\end{align}
The temporal behavior of $\delta T(t)$ is alternatively
understood as impulsive excitation of a damped harmonic
oscillator (DHO).

Writing that $\Omega\equiv\sqrt{\omega_0^2-\Gamma^2}$ and
$\gamma=\sqrt{\Gamma^2-\omega_0^2}=i\Omega$,
the DHO function $f(t)$ can be rewritten as
\begin{align}
 f(t)\propto 
 \begin{cases}
  &e^{-\Gamma t} \dfrac{\sin \Omega\,
  t}{\Omega} \quad (\omega_0>\Gamma: \text{underdamped}) \vspace*{2ex}\\
  &e^{-\Gamma t} \dfrac{\sinh
  \gamma\,t}{\gamma}
  \quad(\Gamma>\omega_0: \text{overdamped})  
 \end{cases}
\end{align}
For the underdamping condition $\omega_0>\Gamma$, $f(t)$ is a
damped oscillation with a period of $2\pi/\Omega$ and damping
rate of $\Gamma$.  
On the other hand, for the overdamping
condition $\Gamma>\omega_0$, $f(t)$ does not oscillate and can be
approximated to be double exponential decays as
\begin{align*}
 f(t)\approx
 \frac{ e^{-\gamma_1 t}-e^{-\gamma_2 t} }{\gamma_2-\gamma_1}, 
\end{align*}
where $\gamma_1\approx \omega_0^2/2\Gamma=D_\text{th}q^2$ and
$\gamma_2\approx 2\Gamma$, with thermal diffusivity
$D_\text{th}=\frac13 \bar{v}_\text{ph}^2 \tau_R$ and
$\Gamma\approx 1/2\tau_R$ (for small value of $q$). This response
appears to rise at a rate $\gamma_2$ and decay at a rate
$\gamma_1$. Thus, in the limit of $\Gamma\gg\omega_0$, we see
that $\gamma_2\gg \gamma_1$, and $f(t)$ leads to an apparent
single exponential decay with virtually a step rise at $t=0$ as
\begin{align*}
 f(t) \approx \exp(-D_\text{th}q^2 t) \quad (t>0),
\end{align*}
which is the well-known thermal-diffusion time response.

The squared Fourier amplitudes (intensity spectra) in frequency
domain for the underdamped and overdamped cases lead respectively
to
\begin{align}
 A^2_\text{ud}(\omega)&\propto
 \frac{1}{(\omega_0^2-\omega^2)^2+4\omega^2\Gamma^2},
 \label{eq:spec} 
\end{align}
and 
\begin{align}
 A^2_\text{od}(\omega)&\propto
 \frac{1}{\omega^2+D_\text{th}q^2}.
\end{align}
These are directly compared with the spectra observed in
spontaneous Brillouin scattering experiments.

\subsection*{Samples}
The samples of SrTiO$_3$ were synthesised single crystals. We
used two samples: Sample A and B. Sample A was purchased at
Shinkosha company, Japan, in 1995, which is very slightly amber
coloured. Sample B was purchased at Crystal Base, Japan, in 2022,
grown by Furuuchi Chemical Corporation, Japan, which was 
non-coloured and transparent. Both Samples A and B were blocks with
dimensions of $10\times8\times7$~mm$^3$, with surfaces oriented
to [001], [110], and [1$\bar 1$0] in the cubic phase.  Our
experiments indicated better thermal conductivity of Sample B
over Sample A (smaller values of $\Gamma$ in Sample B). 

The transient grating was formed along crystalline [001]$_c$
direction, i.e., the wavevector $\vec{q}$ was set parallel to
$[001]_c$.


\subsection*{Experimental setup}

We performed optical heterodyne detection of impulsive stimulated
Brillouin/Thermal scattering, which is essentially identical with
that presented in Ref.\onlinecite{Maznev-OL-1998}, although, instead of
employing a phase mask, we used a spatial light modulator (SLM:
Hamamatsu LCoS-SLM X10468-01) as a zero-order free diffraction
grating for both the pump and probe lasers. The grating pitch,
i.e., the wavelength of the generated excitation, on the SLM
could be quite easily and quickly changed by a personal computer
(PC) since an SLM acts as an external display for a PC. We
prepared several pictures that had several values of the grating
pitches.
We used a pulsed 532~nm radiation of the second harmonic from a
Nd:yttrium-aluminium-garnet laser with a pulse width of 30~ps
(Ekspla PL-1200) as the pump light, and a continuous-wave, single
longitudinal mode of 488~nm radiation from a diode-pumped solid
state laser (Coherent GenesisMX-SLM) as the probe light.  The two
beams are collinearly overwrapped and incident on the SLM, and
are diffracted back toward the sample. The pairs of the 488 and
532~nm beams were collimated and focused simultaneously in the
sample crystal, yielding interference fringes whose pitch exactly
coincides for both the pump and probe lights.
Since the band gap of SrTiO$_3$ is $\approx$3.2~eV, the 532~nm
light is not linearly absorbed, but two-photon absorption is
allowed to produce necessary impulsive heating by non-radiative
relaxation of the electronic excited states as previously
demonstrated in Refs.~\onlinecite{Koreeda2006a} and
\onlinecite{Koreeda2010}. 
%
%
The probe field is diffracted by thermal grating generated by the
pump light, perfectly collinearly propagates with the unperturbed
probe field (local oscillator), and optically-heterodyne-detected
as the beating signal on a fibre-coupled fast photodetector
(10GHz GaAs Amplified Photodetector, Electro-optics Technology:
ET-4000AF).  Thus, the observed signal is proportional to the
diffracted electric field, rather than its intensity, and the
sign of the signal can be positive and negative. Details are
described in Extended Data Fig.1.


\section*{Acknowledgements}

We thank Prof. Toshirou Yagi for his supports and helpful advises
at the very early stage of this research project.
This work was supported by JSPS KAKENHI Grant Numbers
JP17K18765 (A.K. and Y.F.), 
JP19H05618(A.K.), 
JP21H01018(A.K. and Y.F.), 
and JST PRESTO Grant Number JPMJPR10J8 (A.K.).



\section*{Declarations}


\begin{itemize}

 \item Conflict of interest/Competing interests 

       The authors declare no competing interests.

\item Availability of data

      The datasets generated during and/or analysed during this
      study are available from the corresponding author upon
      reasonable request.
      
\item Authors' contributions

      A.K. conceived the experiment. H.O., T.O.
      performed the transient grating measurements, assisted by
      A.K. and Y.F.. H.O and A.K.  analysed and interpreted the
      results.  A.K. wrote the paper. All authors discussed the
      results.

\item Additional information

      Extended Data is available for this paper.

\end{itemize}

\end{document}


\maketitle

\renewcommand{\figurename}{Extended Data Fig.~}


\begin{figure}[h]
 \centering \includegraphics[width=\textwidth]{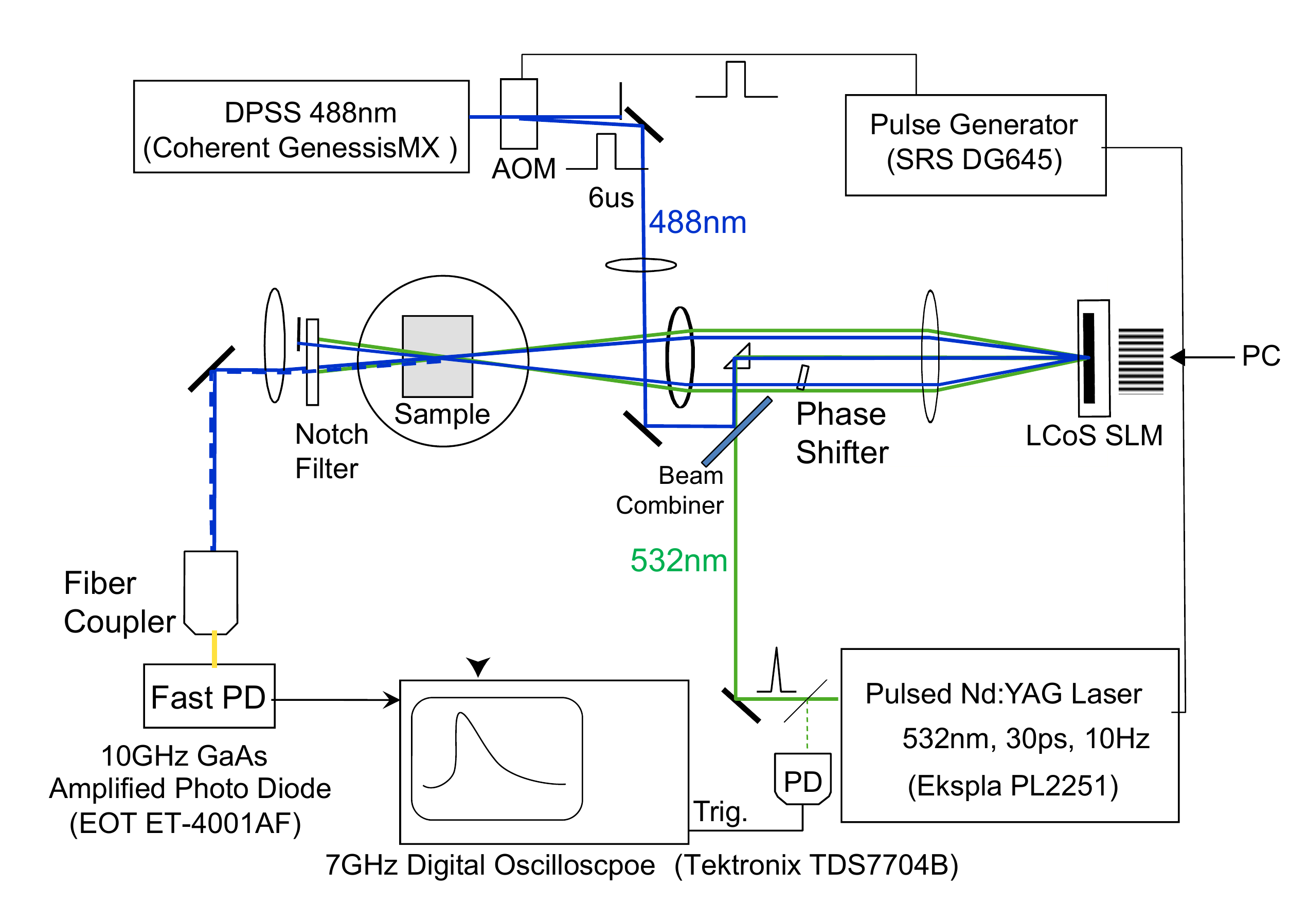}
 \caption{\setlength{\baselineskip}{2.5ex}
 Experimental setup for the generation of coherent thermal
 wave. AOM: acousto-optic modulator, LCoS SLM: liquid crystal on
 silicon spatial light modulator, PD: photodiodes, PC: personal
 computer. We used a pulsed 532~nm radiation of the second
 harmonic from a Nd:yttrium-aluminum-garnet laser with a pulse
 width of 30~ps (Ekspla PL-1200) as the pump light, and a
 continuous-wave, single longitudinal mode of 488~nm radiation
 from a diode-pumped solid state laser (Coherent GenesisMX-SLM)
 as the probe light.  The probe beam was chopped by an AOM to a
 6~$\mu$s pulse to minimize heating of the samples.
 The two beams are collinearly over-wrapped and incident on an
 SLM (Hamamatsu LCoS-SLM X10468-01), and are diffracted back
 toward the sample. The pairs of the 488 and 532~nm beams were
 collimated and focused simultaneously in the sample crystal,
 yielding interference fringes whose pitch exactly coincides for
 both the pump and probe lights. This apparatus is essentially the
 same with that reported in Ref.~24 with the
 phasemask replaced by a PC-controlled SLM.  } \label{fig:setup}
\end{figure}

\newpage

\begin{figure}[t]
 \centering \includegraphics[width=10cm]{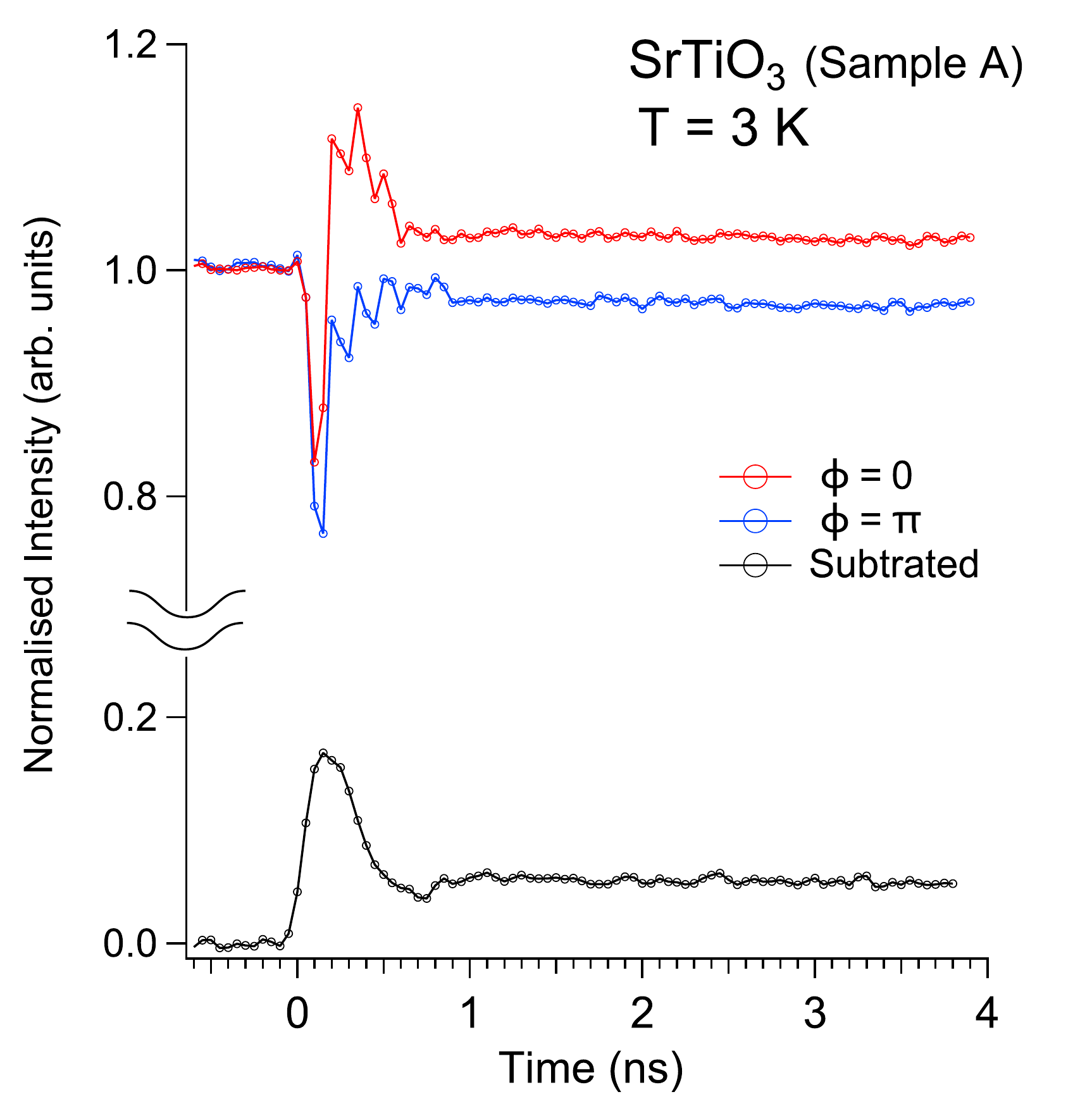}
 \caption{\setlength{\baselineskip}{2.5ex}
 Optically-heterodyne detected signals at phase
 differences 0 and $\pi$, and subtracted signal. The spike and
 ringing responses commonly present in the raw signals were
 eliminated by subtraction procedure and only the response of
 interest can be extracted successfully.
 The probe field is diffracted by the thermal grating generated
by the pump light, perfectly collinearly propagates with the
unperturbed probe field (local oscillator), and
optically-heterodyne-detected on the photodiode as $
I_\text{det}^{\pm}\propto |E_\text{LO}+E^{\pm}_\text{sig}(t)|^2 =
I_\text{LO}^2+I_\text{sig}^2 \pm 2 E_\text{LO}
E_\text{sig}^{\pm}, $ where $I_\text{LO}=|E_\text{LO}|^2$,
$I_\text{LO}=|E_\text{sig}|^2$, and $I_\text{LO}\gg
I_\text{sig}$. The double signs correspond to the relative phases
between $E_\text{LO}$ and $E_\text{sig}(t)$, 0 or $\pi$. Thus,
the observed signal contains a component proportional to the
diffracted electric field $E_\text{sig}(t)$. By acquiring both
signals with relative phases 0 and $\pi$ and subtraction of them
yields $ S(t)=I_\text{det}^+ - I_\text{det}^- \propto 2
E_\text{LO} [E_\text{sig}^+(t)+E_\text{sig}^-(t)], $ This
procedure is particularly advantageous when $E_\text{sig}^{\pm}$
carry common noises or residual signals such as a spike due to
nonlinear electronic response or detector's ringing response.
Since the thermal and acoustic signals change the sign for phase
differences between 0 and $\pi$ while the residual responses not,
i.e., $E_\text{sig}^{\pm}=\textit{Noise} \pm \textit{Signal}$,
the subtraction procedure eliminates the common residual
responses and extracts purely thermal and acoustic responses.
 All traces presented in the main text were obtained
following this procedure.
 }\label{fig:OHD}
\end{figure}

\newpage

\begin{figure}[t]
 \centering
 (a)\\
 \includegraphics[width=8cm]{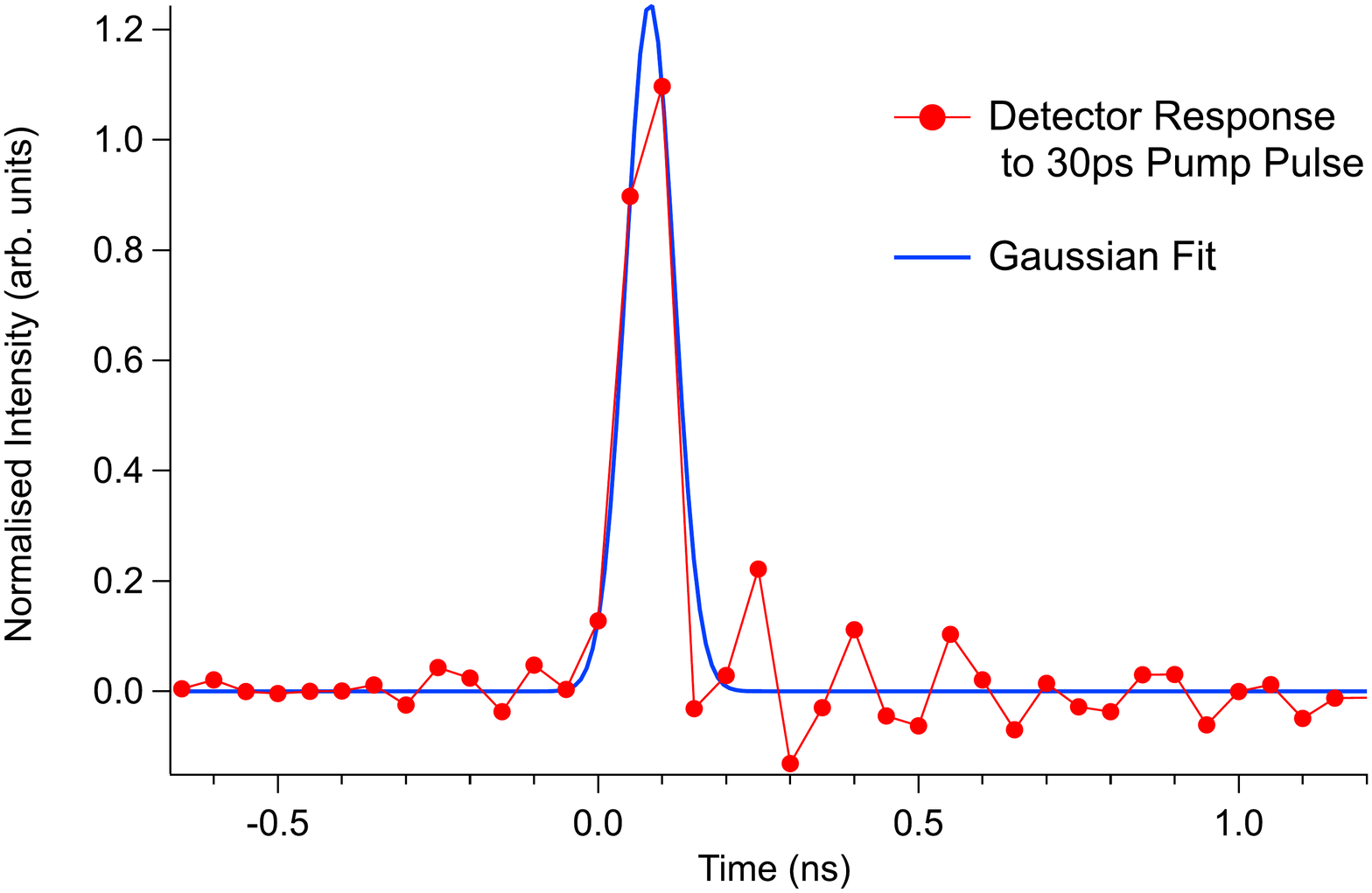}
\vspace*{0.3cm}
 
 (b)\\
 \vspace*{-1cm}
 \includegraphics[width=\textwidth]{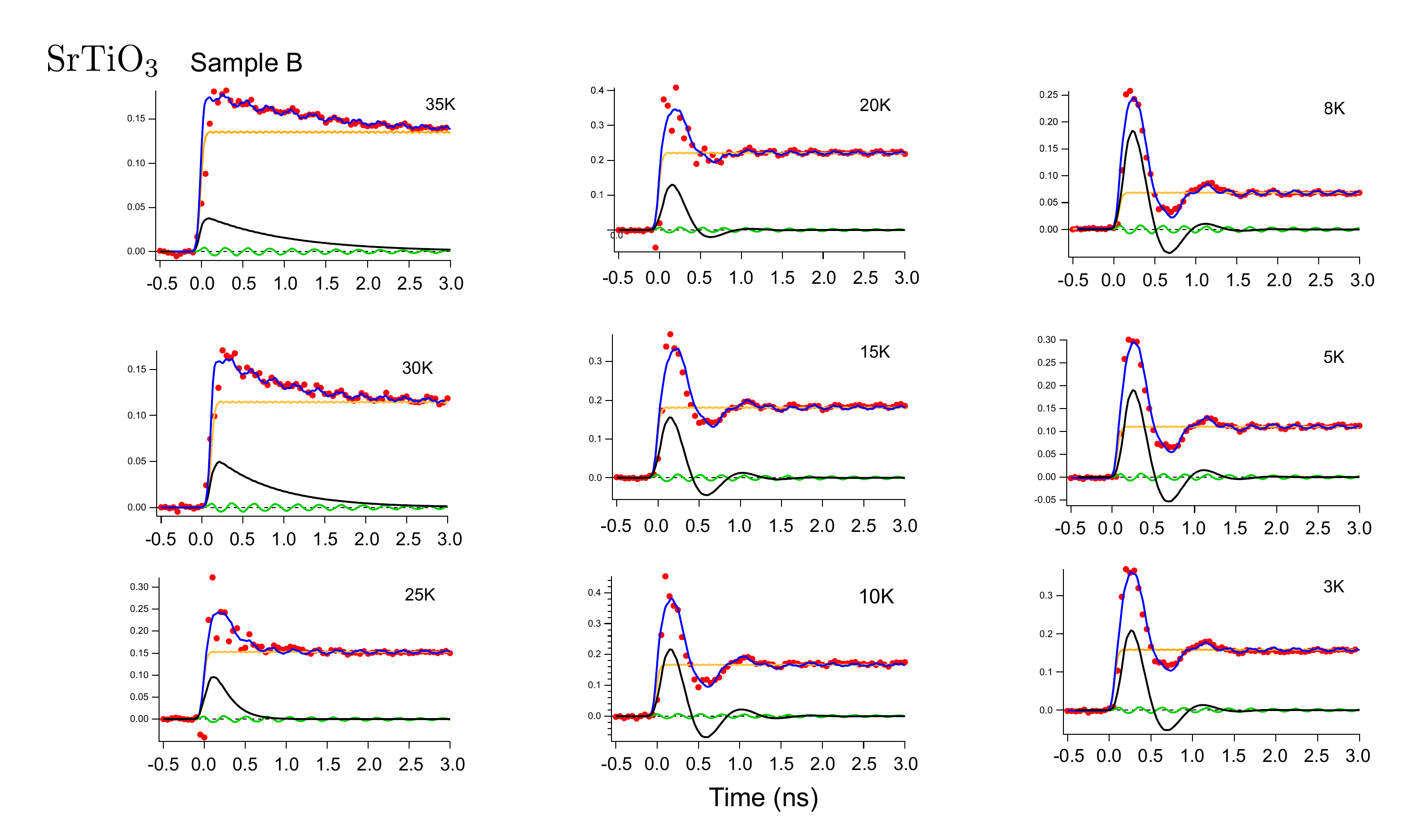}
 \caption{\setlength{\baselineskip}{2.5ex} 
(a) The photodetector response to a 30~ps laser pulse. We
employed a 10~GHz GaAs Amplified Photodetector (Electro-optics
Technology: ET-4000AF).
The blue curve is a gaussian fit to the initial peak. The
effective convolution width was set to 106~ps.  Convolution fit
with raw detector response, exhibiting a 6~GHz ringing, to the
experimental trace was not successful, because the subtracted
signal did not virtually contain such ringing responses owing to
the elimination of the common noise in the subtraction procedure
described in \figurename \ref{fig:OHD}.  On the other hand,
convolution with a gaussian function quite successfully
reproduced the experimental signals. So, we decided to employ a
gaussian profile obtained by fitting to the initial detector
response for convolution fit.
(b) Temperature variation of the observed signals and the fitted
component functions in the convolution fit. The total fitting
function was as follows: $
f(t)=\text{heat}(t)+\text{LA}(t)+\text{step}(t) $.  Here, $
\text{heat}(t)= C_1 \frac{e^{-\Gamma (t-t_0)}}{\Omega}\,\, {\sin[
 \Omega\,\,(t-t_0)]  },\,\, 
 \text{LA}(t) = C_2 \frac{e^{-\Gamma_B
 (t-t_0)}}{\Omega_\text{LA}}\,\, \sin [ 
 \Omega_\text{LA}\,\,(t-t_0) +\phi],\,\,
 \text{step}(t)=\text{const.}\, (t>0),\,\,\text{or}\,\,  0\, (t\leq 0), $
where $\Omega=\sqrt{\omega_0^2-\Gamma^2}$, $\Omega_\text{LA} =
 \sqrt{\omega_\text{LA}^2-\Gamma_\text{LA}^2}$.  $t_0$,
 $\omega_\text{LA}$, $\Gamma_\text{LA}$, and $\phi$ are a time
 offset, LA-phonon frequency, LA damping constant, and relative
 phase of LA mode with respective to second sound (heat
 function), respectively.
 }\label{fig:T-dep_all}
\end{figure}

\newpage

   \begin{figure*}[t]
 \centering \includegraphics[width=0.85\textwidth]{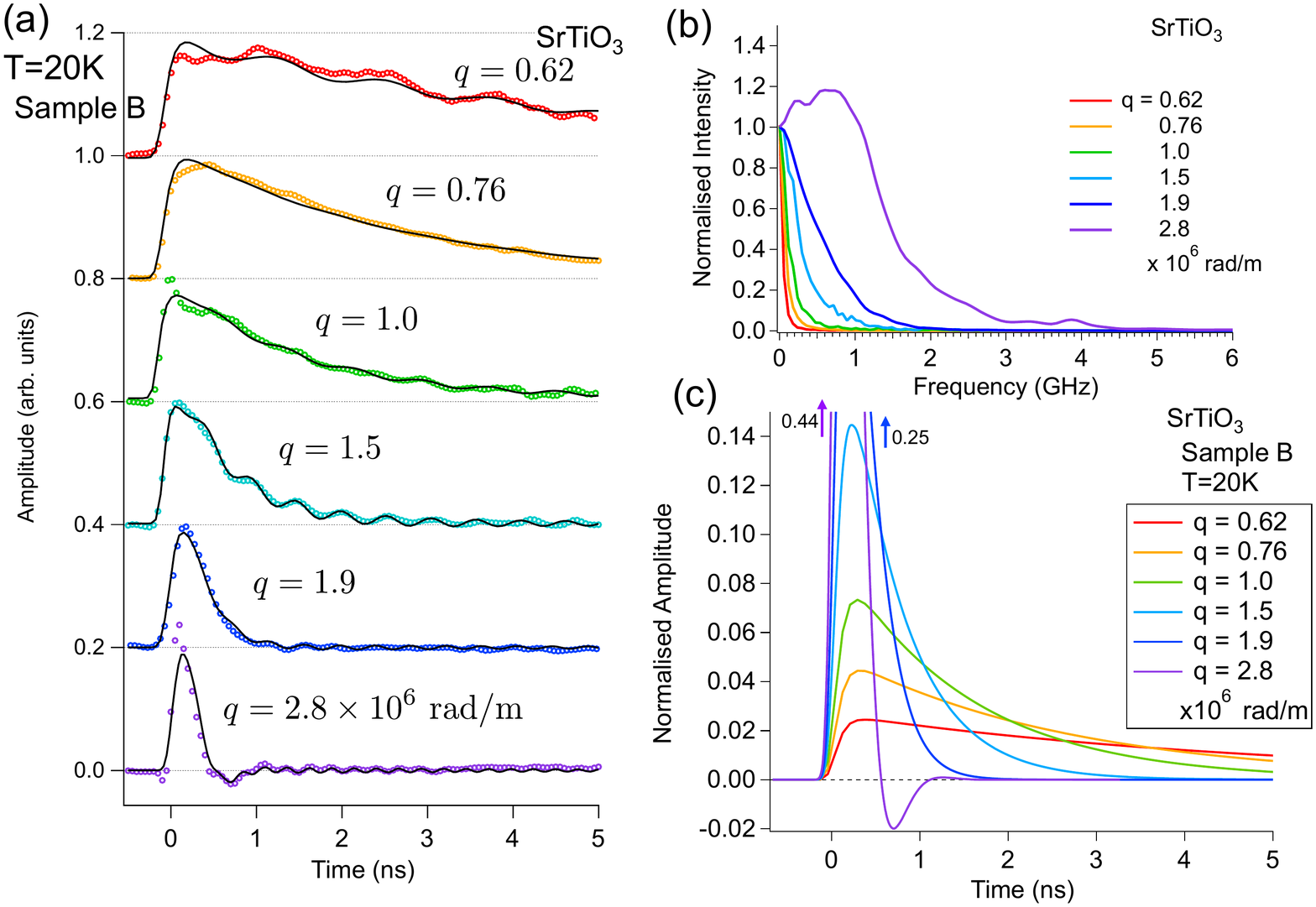}
  \caption{\setlength{\baselineskip}{2.9ex}
(a) Variation of temporal profiles obtained at $T=20$~K for
various wavenumbers ranging from 6.2$\times 10^5$ to $2.8\times
10^6$~rad/m, which correspond to wavelengths ranging from
2.3~$\mu$m to 10~$\mu$m (Sample B). At small wavenumbers, the
signal appears to be an exponential decay, whereas, at the
largest wavenumber, it exhibits an oscillatory behavior.
(b) FFT intensity (squared amplitude) spectra of the temporal
profiles shown in Fig.~\ref{fig:q-dep}a, which serve a direct
comparison with the spectra of spontaneous light-scattering
experiments.  For small wavenumbers, the spectra are narrow
``central peaks'' with a maximum at $\omega/2\pi=0$.
As we increase the wavenumber, the exponential decay becomes
faster and the corresponding central-peak spectrum broadens. At
the largest wavenumber, the time-domain signal is oscillating
with a period of $\approx 1.2$~ns, and, accordingly, the spectrum
exhibits a resonance peak at $\approx 0.8$~GHz.
All the time-domain signals were successfully fitted by the
damped-harmonic oscillator function, with $\Gamma$ being fixed to
a value of $1.25\times 10^{10}$ and 5.56$\times 10^9$~rad/s in
Samples A and B, respectively.
(c) Normalised signal amplitude of the fitted functions, which
    demonstrates the transformation from an exponential decay
    (thermal diffusion) into oscillation (thermal wave) according
    to the wavenumber of the thermal wave.
    }\label{fig:q-dep}
 \end{figure*}

%

\newpage